\documentclass[conference]{IEEEtran}
\IEEEoverridecommandlockouts
\usepackage{cite}
\usepackage{amsmath,amssymb,amsfonts}
\usepackage{algorithmic}
\usepackage{graphicx}
\usepackage{subcaption}
\usepackage{textcomp}
\usepackage{xcolor}
\usepackage{amsmath}
\usepackage{braket}
\usepackage{float}
\usepackage{dblfloatfix}
\usepackage{tikzpagenodes}

\usepackage{fancyhdr}
\fancypagestyle{firstpage}{
   \fancyhf{} 
   \fancyhead[C]{To appear at 2024 Design, Automation and Test in Europe Conference Proceedings, Valencia} 
}
\usepackage[a4paper, total={184mm,239mm}]{geometry}
\def\BibTeX{{\rm B\kern-.05em{\sc i\kern-.025em b}\kern-.08em
    T\kern-.1667em\lower.7ex\hbox{E}\kern-.125emX}}

\begin{document}

\title{Alleviating Barren Plateaus in Parameterized Quantum Machine Learning Circuits: Investigating Advanced Parameter Initialization Strategies\\

}

\author{\IEEEauthorblockN{Muhammad Kashif \IEEEauthorrefmark{1}\IEEEauthorrefmark{2},
Muhammad Rashid\IEEEauthorrefmark{3},
Saif Al-Kuwari\IEEEauthorrefmark{4}, 
Muhammad Shafique\IEEEauthorrefmark{1}\IEEEauthorrefmark{2}}

\IEEEauthorblockA{\IEEEauthorrefmark{1} eBrain Lab, Division of Engineering, New York University Abu Dhabi, PO Box 129188, Abu Dhabi, UAE}
\IEEEauthorblockA{\IEEEauthorrefmark{2} Center for Quantum and Topological Systems, NYUAD Research
Institute, New York University Abu Dhabi}

\IEEEauthorblockA{\IEEEauthorrefmark{3}Department of Computer Engineering, Umm Al Qura University, Makkah, Saudi Arabia}
\IEEEauthorblockA{\IEEEauthorrefmark{4}Qatar Center for Quantum Computing, College of Science and Engineering, Hamad bin Khalifa University, Qatar}
Emails: muhammadkashif@nyuad.edu, mfelahi@uqu.edu.sa, smalkuwari@hbku.edu.qa, muhammad.shafique@nyu.edu
    }

\maketitle

\thispagestyle{firstpage}

\begin{abstract}

Parameterized quantum circuits (PQCs) have emerged as a foundational element in the development and applications of quantum algorithms. However, when initialized with random parameter values, PQCs often exhibit barren plateaus (BP). These plateaus, characterized by vanishing gradients with an increasing number of qubits, hinder optimization in quantum algorithms.
In this paper, we analyze the impact of state-of-the-art parameter
initialization strategies from classical machine learning in random PQCs from the aspect of BP phenomenon. Our investigation encompasses a spectrum of initialization techniques, including random, Xavier (both normal and uniform variants), He, LeCun, and Orthogonal methods. 
Empirical assessment reveals a pronounced reduction in variance
decay of gradients across all these methodologies compared to the randomly
initialized PQCs. Specifically, the Xavier initialization technique outperforms the rest, showing a 62\% improvement in variance decay compared to the random initialization. The He, Lecun, and orthogonal methods also display improvements, with respective enhancements of 32\%, 28\%, and 26\%.
This compellingly suggests that the adoption of these existing
initialization techniques holds the potential to significantly amplify the training
efficacy of Quantum Neural Networks (QNNs), a subclass of PQCs. Demonstrating
this effect, we employ the identified techniques to train QNNs for learning the
identity function, effectively mitigating the adverse effects of BPs. The training performance, ranked from the best to the worst, aligns with the variance decay enhancement as outlined above. 
This paper underscores the role of tailored parameter initialization in mitigating
the BP problem and eventually enhancing the training dynamics of QNNs.
\end{abstract}

\vspace{-0.16cm}
\section{Introduction}
Parameterized quantum circuits (PQCs) are at the forefront of quantum computing research, showcasing the synergy between quantum and other domains. 
PQCs have emerged as a promising candidate to realize and demonstrate the potential quantum advantage in the noisy intermediate-scale quantum (NISQ) era\cite{Preskill:2018}. The NISQ devices are the quantum devices that are available today comprising of limited number of \emph{noisy} qubits with partical error-correction\cite{Bharti:2022}.  
Despite the limitations, the NISQ devices are already achieving the so-called \emph{quantum supremacy}\footnote{The term referring to cases where quantum computers surpass the capabilities of the best classical computers} \cite{Arute:2019,zhong:2020,Wu:2021,Madsen:2022}.

PQCs, consisting of quantum gates with adjustable parameters, serve as a versatile tool for quantum algorithm design. By optimizing these parameters, PQCs can encode complex quantum states or unitary operations, enabling their use in a number of applications such as quantum chemistry simulations \cite{liang:2023,fan:2023}, optimization, quantum error correction\cite{Roffe:2019,Zoratti:2023}, machine learning\cite{Biamonte:2017,Benedetti_2019}. 

Recent findings suggest that PQCs may experience exponentially vanishing gradients as the number of qubits increases. This phenomenon, termed as the \emph{barren plateau} (BP) problem, has been backed by both analytical and numerical evidence \cite{McClean:2018,Cerezo:2021aa}. 
When BP occurs, the variance of parameter gradients vanish exponentially, which results in an optimization landscape that becomes progressively flatter with an increasing number of qubits.
The training of quantum neural networks (QNNs) (or PQCs) becomes challenging when these landscapes are encountered, as the optimization techniques have trouble finding suitable directions to improve the model parameters. 

\begin{figure*}[!htbp]
    \centering
    \begin{subfigure}{0.3\textwidth}
        \includegraphics[height=1.4in]{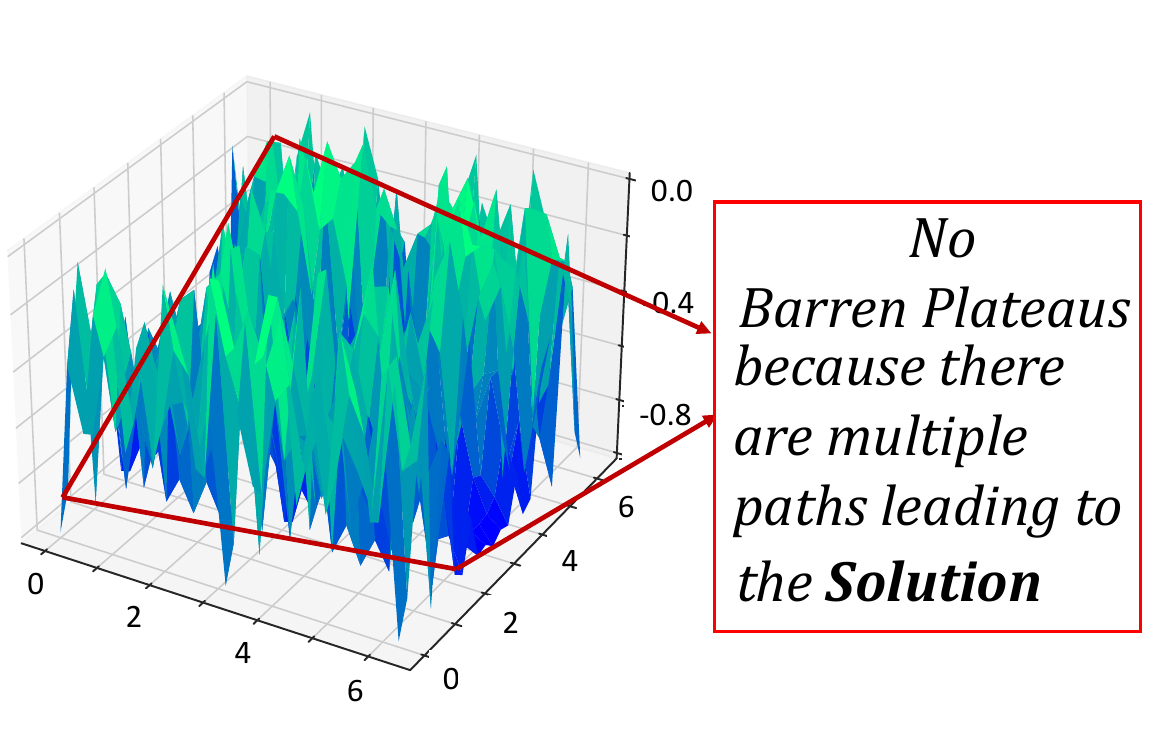}
        \caption{}
        \label{fig:BP2}
    \end{subfigure}
    \hspace{-5pt}
    \begin{subfigure}{0.3\textwidth}
        \includegraphics[height=1.3in]{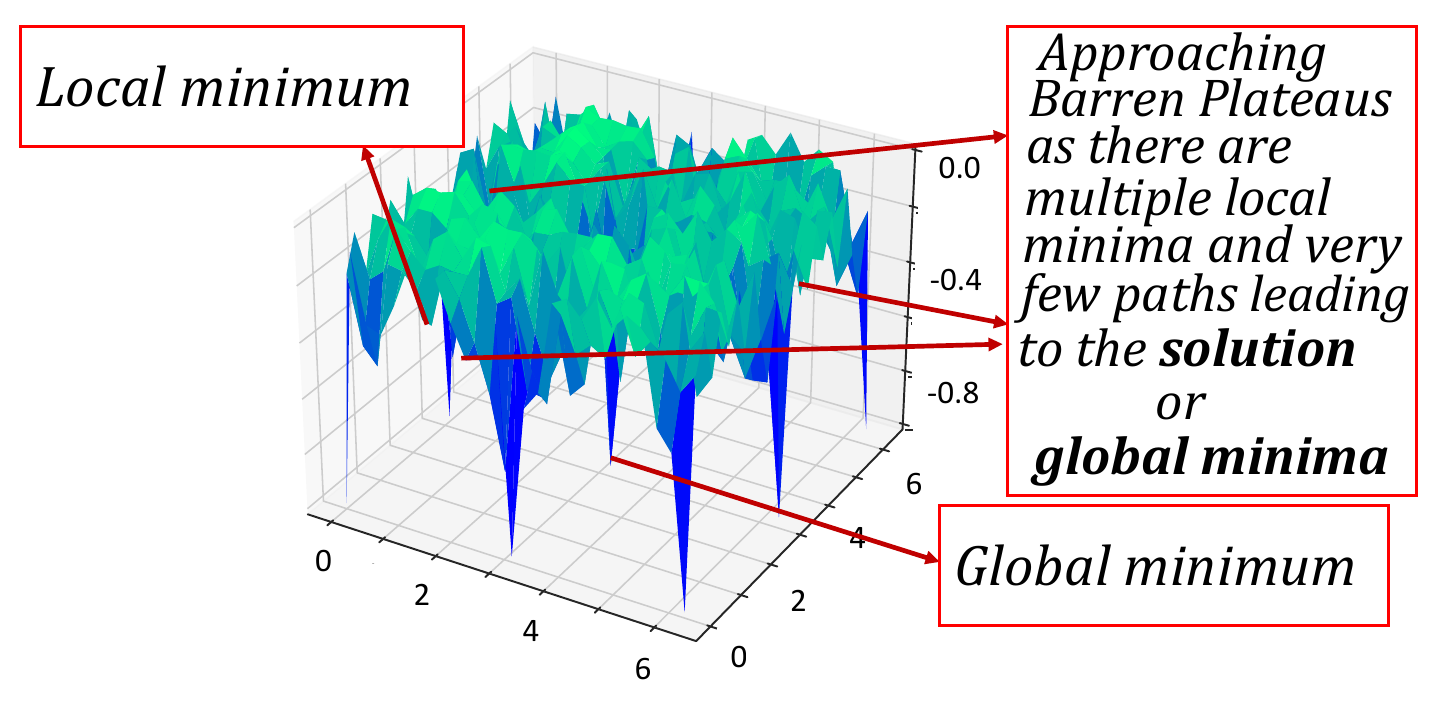}
        \caption{}
        \label{fig:BP5}
    \end{subfigure}
    \qquad \qquad
    \begin{subfigure}{0.3\textwidth}
        \includegraphics[height=1.3in]{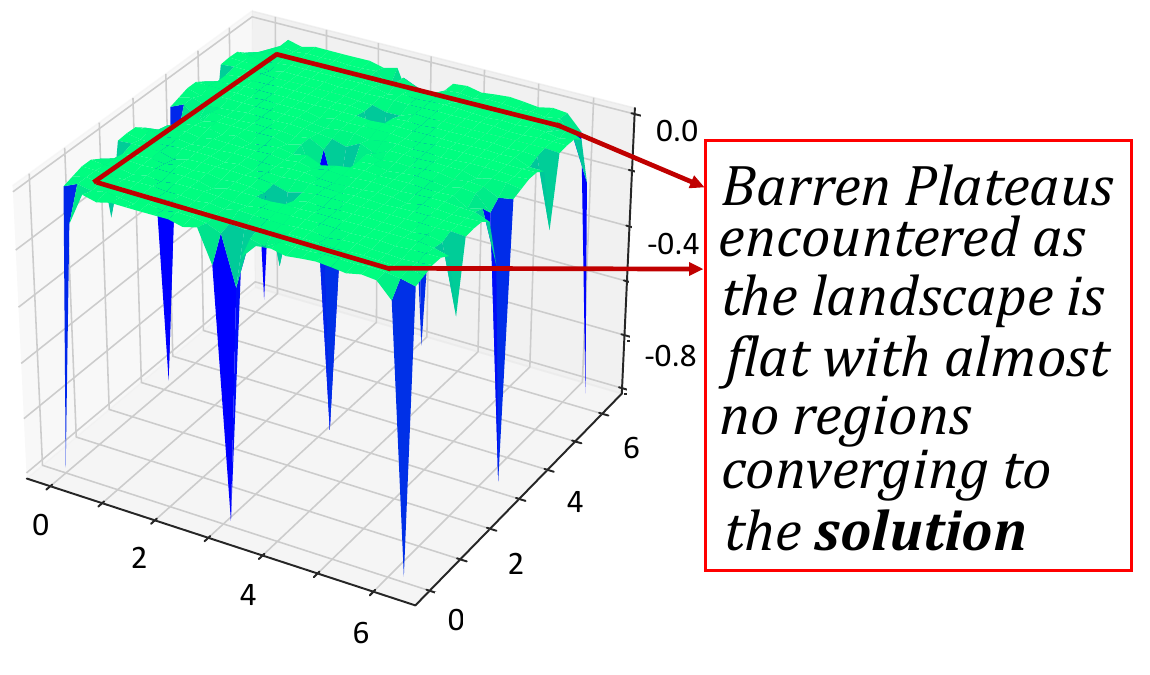}
        \caption{}
        \label{fig:BP10}
    \end{subfigure}
    \vspace{-2pt}
    \caption{Optimization Landscape Demonstrating the Barren Plateaus. (a) 2 qubits (b) 5 qubits (c) 10 qubits }
    \label{fig:BP_demo}
\end{figure*}

\vspace{-1.8pt}
\subsection{Motivational Analysis}
\vspace{-3pt}

A typical illustration of BP phenomenon for various qubits at a constant depth of $100$ layers is depicted in Fig. \ref{fig:BP_demo}. In this context, each layer is composed of two parameterized gates (namely $RX$ and $RY$) for every qubit, along with nearest-neighbor entanglement via the $CZ$ gate. As the number of qubits increases, the optimization landscape tends to flatten, making it harder for the optimizer to converge to a solution.
The BP problem poses a critical challenge because if the quantum circuits are initialized randomly, a wide range of gradient-driven optimization techniques may not be effective. 
Addressing this problem is crucial for the scalability of various algorithms, including QNNs, which is one of most widely explored applications of PQCs\cite{Grant_2018,Chen_2020}.
%
There are various state-of-the-art solutions addressing BPs including block identity encoding\cite{grant:2019}, layer-wise training\cite{Skolik:2021} and residual learning \cite{kashif2023resqnets}.
However, most of these solution are limited because of their implementation complexities and computational demands.


\subsection{Novel Contributions}
\begin{figure}[htbp]
     \centering
     \includegraphics[width=\linewidth]{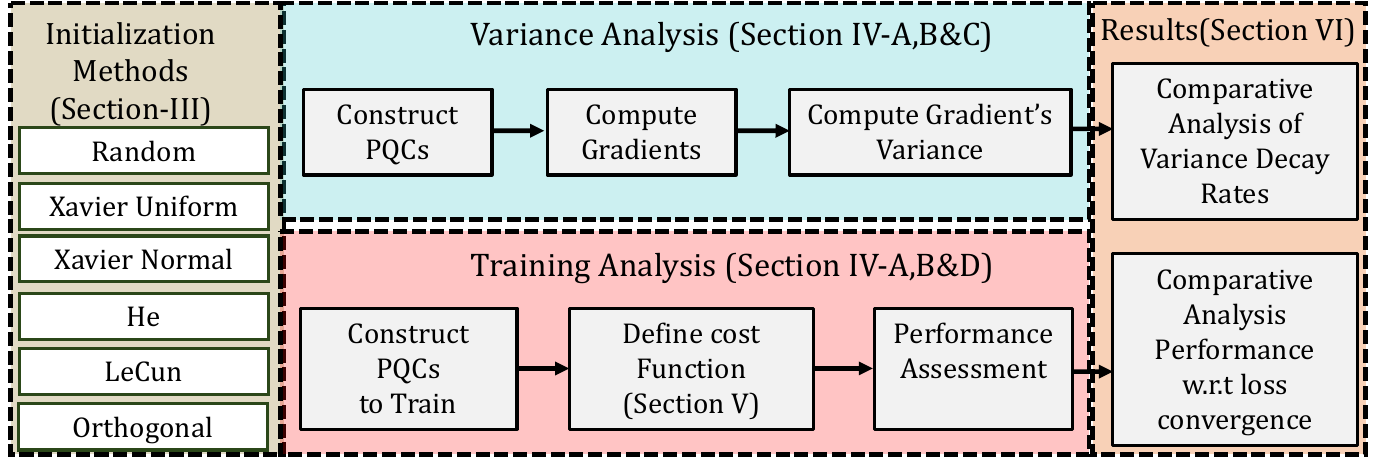}
     \vspace{-2pt}
     \caption{An Overview of the Proposed Contributions.}
     \label{fig:cont}
 \end{figure}

\vspace{-0.25pt}
Below, we highlight our main contributions, a summarized visual representation of which is shown in Fig. \ref{fig:cont}. 

\begin{itemize}
    \item We conducted an extensive analysis, studying the potential influence of classical machine learning initialization methods on the challenging BP phenomenon in PQCs.
    \item Our study spanned a set of well-known initialization techniques from the classical domain, such as: Random, Xavier (covering both its variants), He, LeCun, and the Orthogonal methods.
    \item One of our key findings was the noticeable reduction in gradient variance decay across all techniques when compared with PQCs initialized randomly, a common benchmark.
    \item Our findings indicate the promise these classical initialization methods hold for refining the training processes of a specialized subset of PQCs that is QNNs.
    \item To further support the above claims, we used the evaluated initialization methods to optimize QNNs for a certain problem. The outcomes were encouraging, showcasing notable mitigation of the BP issues and eventually enhancing the learning process.
\end{itemize}

\vspace{-0.2cm}

\vspace{-0.25cm}
Before presenting the details of our work, we provide an overview of some prominent solutions addressing the BP challenge, and highlight their limitations, which are instrumental in understanding the context and significance of our contributions.
 
\vspace{-0.2cm}

\section{Related Work}
\vspace{-0.15cm}
Over the last few years, mitigating BPs in PQCs has been a critical research area due to its significance for the success of various quantum algorithms. Several solutions have been proposed to address barren plateaus:
\paragraph{Block Identity Encoding}
In \cite{grant:2019}, an approach to tackle BPs is proposed, which involves random selection of certain PQC parameters while adjusting others such that the PQC is transformed into a series of shallow blocks, each equivalent to an Identity. By doing so, the approach reduces effective depth and breaks the unitary 2-design structure, facilitating the training process. However, its applicability is contingent on a specific ansatz structure, making broad generalization problematic. Particularly, when one layer of the ansatz has fixed parameters, identifying the other parameters to equate the block to an identity can be computationally challenging.

\paragraph{Quantum Natural Gradient} Another strategy to overcome BP is the use of quantum natural gradient, which considers the unique geometry of quantum state space, potentially leading to better optimization paths\cite{Wierichs:2020}. Despite its promise, the computational cost of the quantum metric tensor, essential for this method, can be high.
\paragraph{Layer-wise Training}
The introduction of layer-wise training, where the depth of the circuit is incrementally increased, optimizing one layer at a time, was proposed in\cite{Skolik:2021} to overcome BPs. By restricting the number of trainable parameters during the initial phases of training, the optimization landscape can become much more favorable for gradient-based optimization.
While this approach benefits from optimizing shorter circuits that are less likely to hit BPs, it could be computationally intensive for very deep circuits. 
\paragraph{Cost Function Locality} In \cite{Cerezo:2021aa}, the dependence of BPs on the cost function is explored, highlighting distinctions between local (measuring a single qubit in multi-qubit systems) and global cost functions (measuring all qubits). While global functions consistently showed BPs regardless of circuit depth, local functions had polynomially vanishing gradients, enabling training upto logarithmic depths. However, \cite{Kashif:2023} later suggested that local cost functions might not be optimal for practical tasks, particularly multiclass classification, to avoid BPs.
\paragraph{Beta Distribution} 
In \cite{Kulshrestha:2022}, a strategy employing beta distribution for parameter initialization and data-driven hyperparameter estimation was introduced. Additionally, it incorporates unique perturbations in each gradient descent step to counteract BPs. However, such hyperparameter estimation risks overfitting and thus may compromise generalization. Moreover, the perturbations might slow down convergence and introduce unexpected dynamics in optimization landscape.

\paragraph{Residual Approach} A recent study suggested incorporating the residual approach in QNNs to address the BPs \cite{kashif2023resqnets}. However, their approach suggest segmenting the standard QNN architecture into multiple quantum nodes with PQCs of different depths. This could result in an increased overhead, particularly in terms of hardware resources and qubit usage which is matter of concern in NISQ devices.


\vspace{-0.15cm}
\section{Parameter Initialization in Neural Networks} \label{sec:params_init}
\vspace{-0.15cm}
The process of selecting initial parameter values for optimizing deep neural networks (DNNs) via gradient-based methods is one of the most influential hyperparameter choices within the realm of deep learning systems. Poor initialization can lead to issues like slow convergence, vanishing or exploding gradients, and suboptimal performance. 
Below, we discuss some notable state-of-the-art initialization approaches proposed to efficiently train DNNs, which we have also used in this paper.

\vspace{-0.18cm}
\subsection{Random Initialization}
\vspace{-0.1cm}
Random initialization is a common method for parameter initialization in DNNs, where parameters are randomly sampled. This prevents symmetry in updates, ensuring neurons in a layer learn diverse features. 
However, such random weights can cause vanishing or exploding gradients, as they are propagated through the network layers, complicating and slowing the training of deep networks.


\vspace{-0.17cm}
\subsection{LeCun Initialization}
\vspace{-0.1cm}
LeCun initialization\cite{LeCun:2012} is a widely-used technique in deep learning for setting DNN weights, ensuring gradients don't vanish or explode. It adjusts initial weights based on the number of input units in a layer. Specifically, for a layer with $n_{in}$ input units and $n_{out}$ output units, the weights are initialized are drawn from a normal distribution of mean 0 and variance:
\vspace{-0.12cm}
$$\text{Var}(W) = \frac{1}{n_{in}}$$
\vspace{-0.12cm}
Or, alternatively, with a uniform distribution between:
\vspace{-0.12cm}
$$-\frac{1}{\sqrt{n_{in}}} \text{ and } \frac{1}{\sqrt{n_{in}}}$$

\vspace{-0.12cm}
\subsection{Xavier Initialization}
\vspace{-0.1cm}
Xavier Glorot and Yoshua Bengio built on LeCun's work, introducing a weight initialization method \cite{glorot:2010} that accounts for both input ($n_{in}$) and output units ($n_{out}$) to enhance the training process. Xavier a.k.a Glorot initialization addresses the vanishing and exploding gradients by ensuring well-scaled initial weights in DNNs. It has two popular variants: Xavier Normal and Xavier Uniform.

\paragraph{Xavier Normal}
The xavier normal initialization sets the weights of a layer with $n_{in}$ input and $n_{out}$ output units according to a Gaussian (normal) distribution with mean $0$ and variance $\sigma^2 = \frac{2}{n_{in}+n_{out}}$. Mathematically, the weights are initialized as;
\vspace{-0.2cm}
$$\mathrm{W} \sim \mathcal{N}(0, \frac{2}{n_{in}+n_{out}})$$
\vspace{-0.3cm}


\paragraph{Xavier Uniform}
In Xavier Uniform initialization, the weights of a layer with $n_{in}$ input and $n_{out}$ output are sampled from a uniform distribution within the range $\mathbf{-limit, limit}$, where limit = $\sqrt{\frac{6}{n_{in}+n_{out}}}$. Mathematically, the weights are initialized as;
\vspace{-0.3cm}
$$\mathrm{W} \sim \mathcal{U}\left(-\sqrt{\frac{6}{n_{in}+n_{out}}}, \sqrt{\frac{6}{n_{in}+n_{out}}} \right)$$

\vspace{-0.2cm}
\subsection{He Initialization}
\vspace{-0.1cm}
He initialization\cite{He:2015}, named after its creator Kaiming He, is another popular initialization technique which also aims to tackle the vanishing gradient problem.
Mathematically, He initialization sets the weights of a layer with $n_{in}$ input units and $n_{out}$ output units by sampling from a Gaussian (normal) distribution with mean 0 and variance variance $\sigma^2 = \frac{2}{n_{in}}$. In other words, the weights  $\mathrm{W}$ are initialized as:
\vspace{-0.2cm}
$$ W \sim \mathcal{N} \left(0, \frac{2}{n_{in}}\right)$$


\vspace{-0.4cm}
\subsection{Orthogonal Initialization}
\vspace{-0.1cm}
Orthogonal initialization, used in deep learning, initializes DNN weights such that they are orthogonal to each other, aiding optimization in deep architectures \cite{hu:2020}. This ensures stable training by preventing drastic changes in weight directions. Mathematically, the weight matrix $\mathcal{W}$ is initialized with an orthogonal or unitary matrix, often achieved through methods like QR decomposition or singular value decomposition (SVD).



\vspace{-0.08cm}
\section{Our Methodology} \label{sec:methodology}
\vspace{-0.15cm}
This paper examines the influence of classical machine learning's state-of-the-art parameter initialization techniques on PQC's behavior, emphasizing BP from the analytical perspective. 
We consider hardware-efficient ansatz (HEA), which is a versatile quantum circuit structure widely used in NISQ applications. HEA consists of multiple layers, each containing single and two-qubit gates. HEAs are both implementable and simulatable on current quantum hardware, making them a popular choice for various quantum algorithms today. The HEA we have used can be described by the following equation. 
\vspace{-0.2cm}
\begin{equation}\label{eq1}
    U(\theta) = \prod_{i=1}^K U_{\text{entangle}}U_{\text{rotation}}(\theta_i)
\end{equation}

\begin{figure*}[!t]
    \centering
    \hspace{0.25cm}
    \includegraphics[scale=0.47]{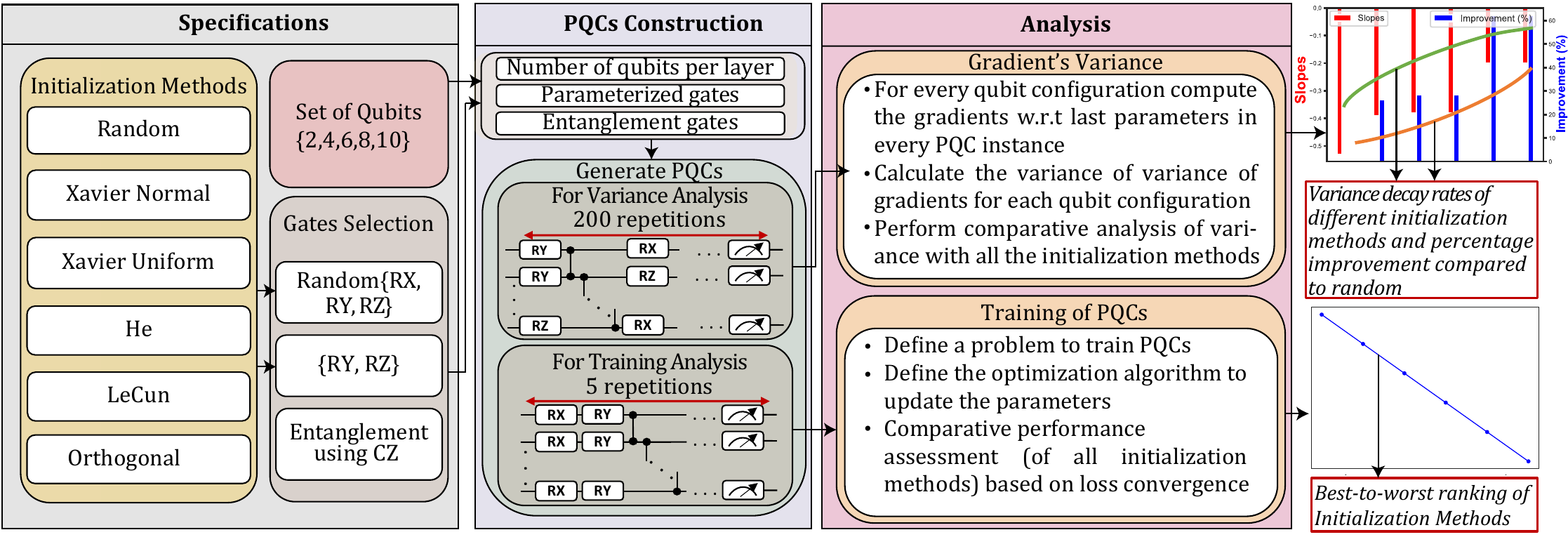}
    \caption{An Overview of Key Steps For Variance and Training Analysis of PQCs via Different Initialization Methods. We consider parameters generated from different initialization approaches for PQCs of different widths (number of qubits). $200$ PQCs are constructed for variance analysis of gradients, where parameterized gates for each PQC are randomly selected and qubit count. The PQCs of depth $5$ (each qubit has two parameterized gates), and width $10$ are constructed for training analysis with different initialization approaches.}
    \label{fig:method_complete}
\end{figure*}

\vspace{-0.2cm}

where $U_{\text{entangled}}$ is a two-qubit gate, typically $CZ$ to entangle two qubits and $U_{\text{\text{rotation}}}(\theta_i)$ is a single qubit parameterized gate and $K$ is the total number of repetitions. 
Adhering to the standardized definition of BP we maintain a consistent approach by maintaining substantial depth in PQCs while progressively increasing the number of qubits. 
Our methodology can broadly be divided into two major steps; (1) Examining the variance of gradients in random PQCs with different initializations and (2) Training PQCs for specific problem to assess potential advantages from a learning perspective. A comprehensive view of our methodology is presented in Fig. \ref{fig:method_complete}.

\vspace{-0.1cm}
\subsection{Specifications}
\vspace{-0.1cm}
Before commencing with experimentation, it is imperative to define specific design considerations. These are outlined as:

\textbf{Qubit Specification.} Selection of qubits for PQC design is crucial. In our study, we consider a set $Q={2,4,6,8,10}$

\textbf{Gates Selection.} The selection of single and two-qubit gates for PQC construction is pivotal. For the variance analysis of gradients, we utilize a gate randomly drawn from $G = \{RX, RY, RZ\}$. In contrast, for training analysis, every qubit is subjected to two distinct parameterized gates, namely $RX$ and $RY$. Furthermore, it is essential to designate a gate for qubit entanglement. We opt for the controlled phase (CZ) gate to entangle neighboring qubits, denoted as:
$E = \prod_{j=1}^{q-1} \text{CZ}_{j, j+1},$
where $q$ represents the qubit count.

\textbf{Parameter Initializations.} Finalizing the approach for parameter initialization of the selected one-qubit gates is essential. In this study, we employ a series of initialization techniques, captured by the set 
$T = \{\text{Random, X-Normal, X-Uniform, He, LeCun, Orthogonal}\}.$

\subsection{PQCs Construction}
Upon finalizing the set of specifications, we independently construct PQCs for both variance and training analyses. This distinction arises from the necessity to attain a substantial depth in PQCs to observe variance decay. Conversely, training these deep PQCs is computationally demanding, prompting us to limit the depths for the sake of efficient experimentation.


\paragraph{PQCs For Variance Analysis}
This step involves generating random PQCs and evaluating them across multiple trials for different qubit configuration specified in previous step. 
For each  $q \in Q$, we construct $200$ random PQCs of the form, as shown in Eq. \ref{eq1}.
\vspace{-0.6cm}
    \begin{equation}\label{eq:var_PQC}
        U(\theta) = \prod_{i=1}^K U_{q,i,t}(\theta, E)
    \end{equation}
\vspace{-0.1cm}
where $q$ is the number of qubits, $i$ ranges from 1 to 200, $\theta$ is the set of parameters (for randomly selected parameterized gate) initialized using techniques in $T$ and $E$ represents the nearest neighbor qubit entanglement operation for $q$ qubits.

\paragraph{PQCs for Training Analysis}
In this step we construct PQCs for training on a particular problem (see Section \ref{sec:exp_setup} for details). As mentioned earlier, these PQCs are slightly different than the ones for variance analysis. For training analysis we construct PQCs of the form:
\vspace{-0.18cm}
\begin{equation}\label{eq:train_PQC}
    U(\theta) = \prod_{l=1}^{L}\left(\prod_{j=1}^{n}RY(\theta)RX(\theta).\prod_{k=1}^{n-1}\left(CZ_{k,k+1}\right)\right)
\end{equation}

\vspace{-0.25pt}
where $L$ is the number of layers/repetitions of PQC and $n$ is the number of qubits.
\subsection{Variance Analysis}  
\vspace{-1.5pt}

We compute the gradients of PQCs defined in Eq. \ref{eq:var_PQC} with respect to their parameters and focus on the variance of these gradient values as a metric for assessing the circuit's behavior and optimization landscape. 
This process is done for all the initializaiton techniques used in this paper. By examining how these initialization methods impact gradient variances, our study provides valuable insights into the role of initialization in quantum circuit training. We now present the work flow of gradients and their variance computation.

\textbf{Gradient Calculation.} 

    For each $U_{q,i,t}(\theta_{q,i,t}, E)$, calculate the gradient $\nabla \theta$ with respect to the last parameter:
    \vspace{-0.3cm}
    $$g_{q,i,t} = \frac{\partial U_{q,i,t}(\theta, E)}{\partial \theta_{\text{last}}}$$
    
    where $g_{q,i,t}$ is the gradient value for the $i^{th}$ PQC of qubit $q$ using initialization $t$ and $\theta_{\text{last}}$ is the last parameter in $\theta$.

\textbf{Storage of Gradient Values.} 
    For each $q \in Q$ and each initialization technique $t \in T$:
    \vspace{-0.3cm}
    $$G_{q,t} = \{ g_{q,1,t}, g_{q,2,t}, \dots, g_{q,200,t} \} $$

\textbf{Variance Calculation.} 
    For each $q \in Q$ and each initialization technique $t \in T$:
    $V_{q,t} = \text{Var}(G_{q,t}) $   where $\text{Var}$ represents the variance calculation.

\subsection{Training Analysis} \label{sec:method_train}
\vspace{-1.5pt}
We now train the PQCs decribed in Eq. \ref{eq:train_PQC} for a specific problem, holding the qubit count and quantum layer depth constant for all the aforementioned initialization strategies, and comparatively assess the training performance.
We use two types of gates per qubit: the $RX$ and $RY$ gates. Additionally, we implement entanglement between nearest-neighbor qubits using $CZ$ gate to facilitate quantum correlations.
The width of our quantum layers, represented by the number of qubits ($n$) in Eq. \ref{eq:train_PQC} is $10$. We maintain a consistent PQC depth of $L=5$ layers. Given this design, the total gate count for the ansatz utilized in the training  is $145$ with parameters count of $100$.

Subsequently, the PQCs, characterized by parameter initialization method, are trained for a specific problem, the details of which are presented in section \ref{sec:exp_setup}. Every training cycle, regardless of the parameter initialization approach, undergoes a fixed number of iterations. The training performance is then evaluated based on the loss function convergence.


\vspace{-0.15cm}
\section{Experimental Setup}\label{sec:exp_setup}
\begin{figure}[htbp]
    \centering
    \includegraphics[width=.85\linewidth]{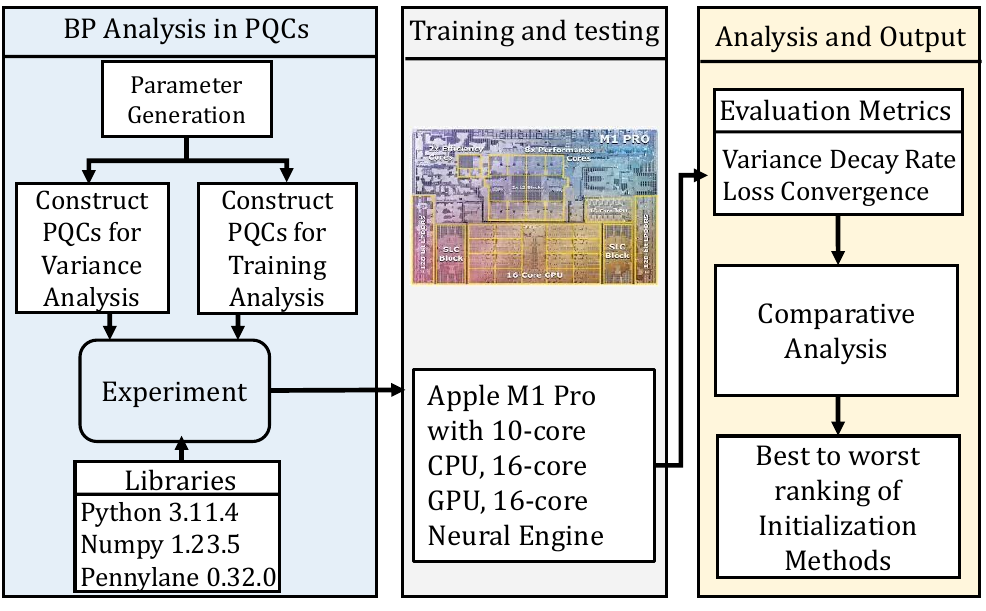}
    \vspace{-5pt}
    \caption{Overview of Our Experimental Tool-flow}
    \label{fig:exp_workflow}
\end{figure}

\vspace{-1pt}
An overview of our experimental tool-flow is depicted in Fig. \ref{fig:exp_workflow}. 
We train the PQCs using all the aforementioned parameter initialization techniques for learning the identity function, which can be described by the following equation.
\vspace{-0.21cm}
$$  C = \bra{\psi(\theta)} (I-\ket{0}\bra{0}) \ket{\psi(\theta)} = 1- p_{\ket{0}} 
$$
We adopt a global cost function framework, wherein we take measurements from every qubit in the network. Hence, the aforementioned cost function restated as:
\vspace{-0.15pt}
\begin{equation}\label{eq:CF_eq}
    C = \bra{\psi(\theta)} (I-\ket{00\ldots0}\bra{00\ldots0}) \ket{\psi(\theta)} = 1- p_{\ket{00\ldots0}} 
\end{equation}
\vspace{-0.45pt}
The PQCs are subjected to a maximum of $50$ training iterations. To optimize our cost function, we use two different optimizers: \emph{Gradient Descent} and \emph{Adam}, both with a step size of $0.1$. The experiments are performed using Pennylane, a versatile python library dedicated to differentiable programming on quantum computers \cite{Bergholm:2018}.


\vspace{-2pt}
\section{Results and Discussion} \label{sec:results}

We now present the findings derived from our experiments, systematically arranged and analyzed to draw significant conclusions. 
The comparative analysis between various parameter initialization methods serve to elucidate the nuances in outcomes, providing a comprehensive understanding of the scope and implications of this paper. 
\begin{figure*}[htbp]
    \begin{subfigure}{0.3\textwidth}
        \includegraphics[scale=0.31]{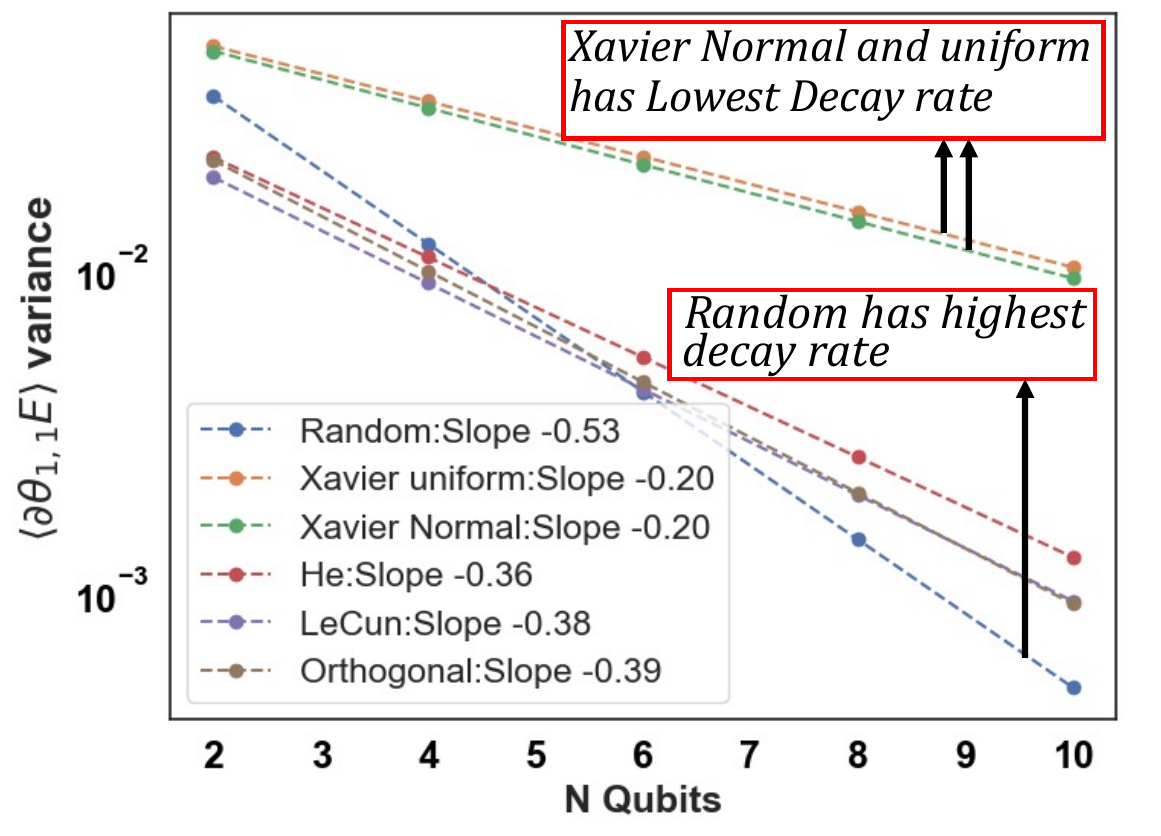}
        \caption{}
        \label{fig:res_var}
    \end{subfigure}
    \qquad
    \begin{subfigure}{0.3\textwidth}
        \includegraphics[scale=0.23]{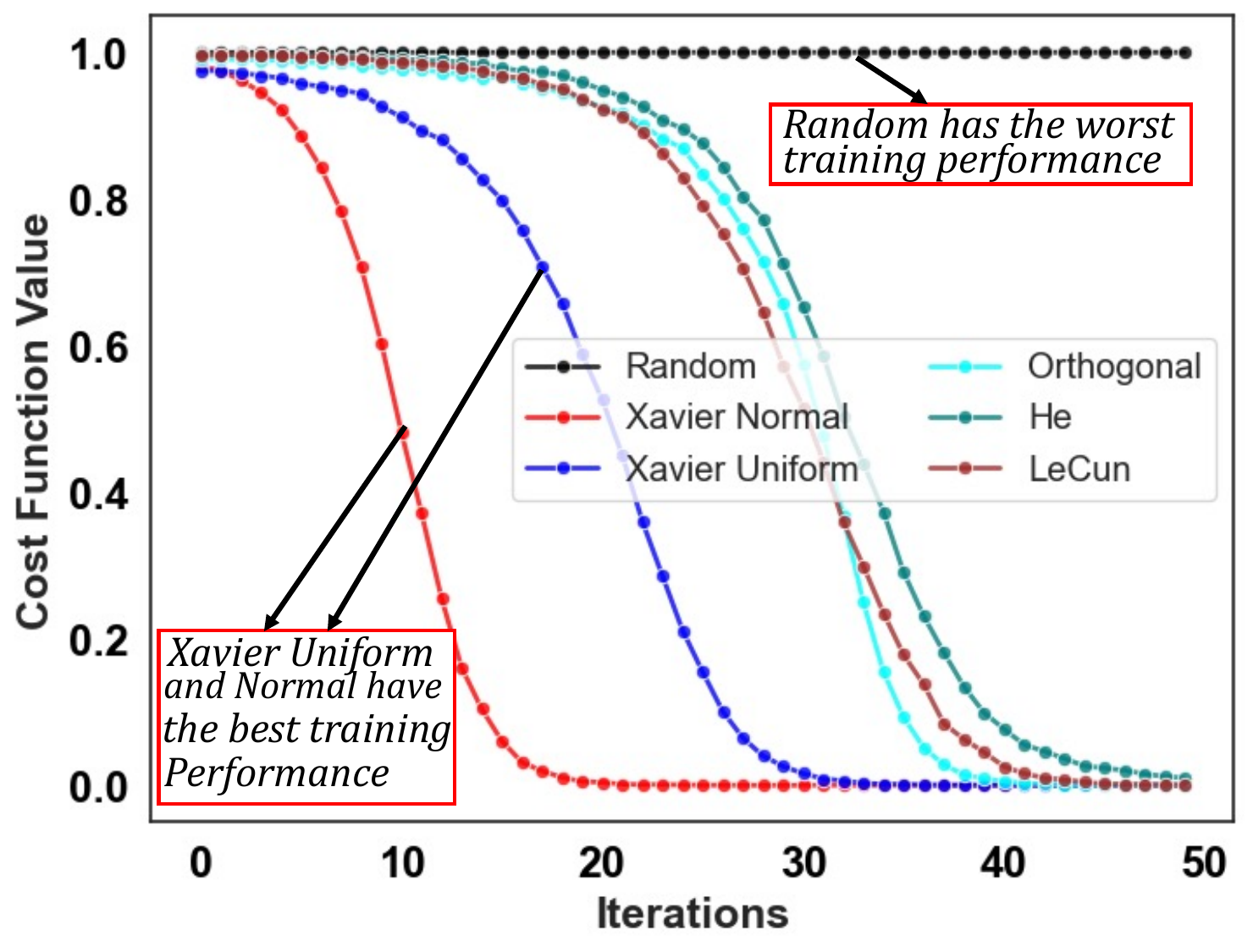}
        \caption{}
        \label{fig:res_train_GD}
    \end{subfigure}
    \qquad
    \begin{subfigure}{0.3\textwidth}
        \includegraphics[scale=0.23]{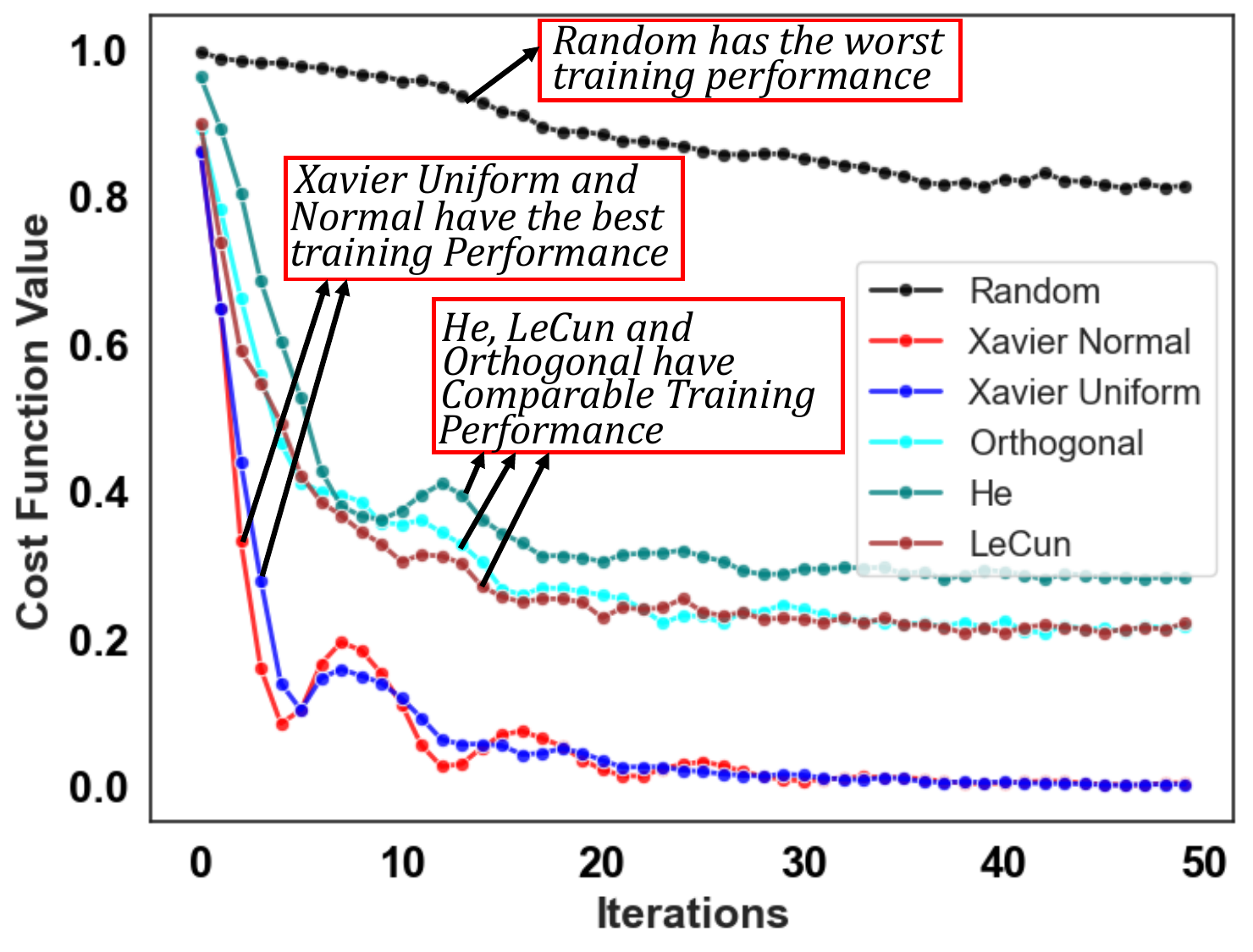}
        \caption{}
        \label{fig:res_train_adam}
    \end{subfigure}
    \vspace{-5pt}
    \caption{Results of Variance and Training Analysis. (a) Variance Decay Rate (b) Loss Convergence with Gradient Descent Optimizer For 10 Qubit Layers (c) Loss Convergence with Adam Optimizer For 10 Qubit Layers}
    \label{fig:Results}
\end{figure*}

\vspace{-0.1pt}
\subsection{Variance Analysis}
\vspace{-0.25pt}
In our evaluation of the variance of gradients across various initialization methods, we utilized a diverse set of 200 randomly generated PQCs. The empirical results obtained are presented in Fig. \ref{fig:res_var}. Our observations lead to several salient points worthy of a detailed discussion:

\textbf{Random Initialization:} 
The variance decay rate of randomly initialized PQCs has a notably steeper negative slope compared to other initialization methods. This suggests a higher susceptibility to encounter BP in the cost function landscape, impacting both convergence rate and training stability.

\textbf{Xavier Initialization:} Both Xavier normal and Xavier uniform techniques emerge as notably superior in our comparisons. The variance decay rates associated with these methods demonstrated a substantial improvement, with an enhancement of approximately 62.3\% over random initialization approach. This highlights the effectiveness of Xavier initialization in maintaining gradient information, potentially avoiding BPs and ensuring faster, stable QNN convergence.

\textbf{He, LeCun, and Orthogonal Initialization:}
All these initialization techniques show a very similar performance. 
However, He initialization exhibited the highest improvement in variance decay rate among these three, with a 32\% improvement when compared with the random initialization. This suggests a potential synergy between the properties of He initialization and characteristics of  PQCs, which eventually can lead to better learning performance. 

LeCun initiazation method known for its adaptability in conventional DNNs, exhibited a 28.3\% improvement in variance decay rate over random initialization in the context of PQCs, which is a significant improvement and can lead to a better training performance compared to random initialization.

The orthogonal initialization method also demonstrated a notable improvement in performance, achieving a variance decay rate that was improved by 26.4\% compared to the random initialization. The essence of the orthogonal approach, centered around maintaining the magnitude of gradients throughout layers, seems to offer tangible benefits in the context of PQCs. 

\vspace{-2pt}

\subsection{Training Analysis}
\vspace{-2.5pt}
In the previous section, we highlighted noticeable enhancements in the gradient variance decay rate across the different initialization methods. An essential inquiry that naturally arises is whether these enhancements in the variance decay rate are mirrored during the training of PQCs.

To investigate further, we trained the PQCs as described in Section \ref{sec:method_train}, for the problem defined in Eq. \ref{eq:CF_eq}. 
The parameters for these PQCs were initialized leveraging all the initialization methodologies. The parameters optimization was performed using two prominent algorithms: Gradient Descent and Adam. The training results are presented in Fig. \ref{fig:res_train_GD} and \ref{fig:res_train_adam} respectively for gradient descent and adam optimizer.

The training results typically reflect the earlier findings on gradient's variance decay rates.
Both variants of the Xavier initialization clearly outperformed the other initialization methods. This was evident in their swift convergence, providing a clear testimony to their effectiveness in circumventing the BP problem.
Randomly initialized PQCs seemed to face significant challenges. They quickly became trapped in flat regions of the cost function landscape, leading to a little or no learning.

The He, LeCun, and Orthogonal also performed consistently with their variance decay rates.
These methods exhibited a closely contested race in terms of loss function minimization. While all initialization methods eventually reached the solution for our simple target problem, the convergence rates of these three were notably slower than the Xavier initialization.

Our training analysis highlights the significance of appropriate parameter initialization in PQCs. While techniques like Xavier initialization stand out, others reveal the subtle effects different strategies have on quantum circuit training dynamics.

\vspace{-0.2cm}
\section{Conclusion} \label{sec:conclusion}
\vspace{-0.2cm}
In this paper, we have conducted an investigation into the impact of various state-of-the-art parameter initialization strategies w.r.t barren plateaus (BP) in Parameterized Quantum Circuits (PQCs). Our analysis spanned a range of initialization techniques, namely random, Xavier (normal and uniform), He, LeCun, and Orthogonal methods. 
The primary focus was on assessing the extent of variance decay exhibited by these techniques in comparison to the PQCs initialized randomly.

A significant finding from our empirical analysis was the marked decrease in variance decay across all tested initialization methods when compared to the randomly initialized PQCs. This observation underlines the promise of these refined initialization strategies in potentially amplifying the training efficiency of Quantum Neural Networks (QNNs). 
To validate this proposition, we train QNNs with the objective of learning the identity function. The results affirmed a considerable reduction in the barren plateau challenges when harnessing the aforementioned initialization techniques.

In conclusion, our study underscores the pivotal role played by parameter initialization in shaping the training landscape of QNNs. The empirical evidence presented here contributes to a better understanding of how specific initialization techniques can effectively alleviate the challenges posed by BPs, paving the way for more robust and efficient QNN training.

\vspace{-0.1cm}
\section*{Acknowledgements}
\vspace{-0.14cm}
This work was supported in part by the NYUAD Center for Quantum and
Topological Systems (CQTS), funded by Tamkeen under the NYUAD Research
Institute grant CG008.

\vspace{-0.15cm}
\bibliographystyle{ieeetr}
\bibliography{main.bib}

\end{document}